\documentclass[12pt,preprint]{aastex}
\usepackage[footnotesize]{caption2}

\usepackage{fancyhdr}

\begin{document}\small
\def\gapprox{\mathrel{\vcenter{\offinterlineskip \hbox{$>$}
    \kern 0.3ex \hbox{$\sim$}}}}
\def\lapprox{\mathrel{\vcenter{\offinterlineskip \hbox{$<$}
    \kern 0.3ex \hbox{$\sim$}}}}

\newcommand{\Dt}[0]{\bigtriangleup t}
\newcommand{\Dx}[0]{\bigtriangleup x}
\newcommand{\E}{\mathcal{E}}
\newcommand{\D}{\bigtriangleup}
\newcommand{\beq}{\begin{equation}}
\newcommand{\eeq}{\end{equation}}
\newcommand{\mm}[2]{\textrm{minmod}\left({#1},{#2}\right)}
\newcommand{\sign}{\textrm{sign}}
\newcommand{\nf}{\mathcal{F}}

\title{Forced field extrapolation of the magnetic structure of the H$\alpha$ fibrils in solar chromosphere}

\author{Zhu Xiaoshuai, Wang Huaning, Du Zhanle and He Han}

\affil{Key Laboratory of Solar Activity, National Astronomical Observatories, Chinese Academy of Sciences, Beijing 100012, China; xszhu@bao.ac.cn}

\begin{abstract}
We present a careful assess of the forced field extrapolation using Solar Dynamics Observatory/Helioseismic and Magnetic Imager (SDO/HMI) magnetogram. The convergence property is checked by several metrics. The extrapolated field lines below 3600km appear to be aligned with most H$\alpha$ fibrils observed by New Vacuum Solar Telescope (NVST). In the region where magnetic energy far larger than potential energy, field lines computed by forced field extrapolation still consistent with the patterns of H$\alpha$ fibrils while non-linear force free field (NLFFF) results show large misalignment. The horizontal average of lorentz force ratio shows the forced region where force-free assumption is failed can reach the height of $1400-1800km$. The non-force-free state of the chromosphere is also confirmed by recent radiation magnetohydrodynamics (MHD) simulation.
\end{abstract}

\keywords{Magnetic fields - Magnetohydrodynamics (MHD) - Methods: numerical - Sun: chromosphere}

\section{Introduction}

It is generally believed that H$\alpha$ fibrils align with magnetic field in chromosphere, due to the reason that ``frozen in field'' effect only allow the motion of fibril plasma along the magnetic field lines. \citet{rs11} compared the orientation of fibrils and magnetic field obtained via high resolution spectropolarimetric observations of Ca II lines. They found most fibrils are aligned with the field lines, while a few cases showed large misalignment. They speculated the time and height difference of Ca II line core and Stokes Q and U peak might explain the misalignment. \citet{spl13} also analyzed the observation of He I triplet, the projected angle of fibrils and the field lines align within an error of $\pm 10^{\circ}$.

Based on the characteristic of alignment, \citet{wts08} extend the classical preprocessing routine of magnetogram by additional consideration of minimizing the angle between the horizontal projection of field lines and fibrils. \citet{jyr11} quantitatively assessed the non-potentiality of fibrils by the shearing angle between the orientation of fibrils and potential field lines extrapolated from longitudinal component of magnetogram.

As the main tool to compute the magnetic field above the photosphere, extrapolation of the magnetogram can be used to compare with the fibrils directly. Although NLFFF models successfully reconstructed many magnetic structures of filaments, EUV bright points and active regions in the corona \citep{ydk01,b04,hwy11,shl12,fwh12}, \citet{mds08} found small structures and field connectivity in a sun-like reference model \citep{b04} can not be well reproduced with forced boundary of photosphere used as input. Furthermore, extrapolation models are much less tested in chromosphere. \citet{wys00} and \citet{wys01} calculated the magnetic field of AR 7321 by boundary integral equation \citep{ys97,ys00}. They found, under modest spatial resolution of $2''$, the field lines are mostly in agreement with fibrils as well as some misalignments.

\citet{zwd13} presented a new implementation of MHD relaxation method to extrapolate the magnetic field on the sun. In this method, a high $\beta$ region like the photosphere near the bottom boundary is introduced by setting a sun-like atmosphere model as the initial condition. By using ``stress and relax'' technique, the MHD system finally reaches an equilibrium state in which the forced magnetic field is balanced by gravity and pressure gradient. No preprocessing is required since a forced boundary is consistent with the extrapolation. The method has been tested with a sun-like reference field. The reference field is the final state of a flux rope emergence model designed by \citet{mgd04}. Detailed diagnostics showed that, compared with the optimization method \citep{w04}, the forced field extrapolation can reconstruct the magnetic field closer to the reference field.

In this paper, we report the modifications of the forced field extrapolation with respect to the version of \citet{zwd13} and its application to reconstruct the magnetic field in chromosphere of active regions (AR) 11967 with SDO/HMI \citep{ssb12} magnetograms used as input. Only the magnetic field in chromosphere is considered due to the reasons that chromosphere is a more forced region than corona and the spatial resolution of H$\alpha$ image can reach $0''.16$ \citep{lxg14}. We will assess the extrapolation with the convergence, different forces and resemblance to H$\alpha$ fibrils observed by NVST \citep{lxg14}. We are also interested with the difference between the results of forced field and NLFFF extrapolation.

The paper is organized as follows: the modifications of forced field extrapolation method are described in section 2; the extrapolation results of AR 11967 are accessed detailed in section 3; the data used here are briefly described in section 4; we discuss in section 5 and finally conclude in Section 6.

\section{Method}

We extrapolate the magnetic field by solving the MHD equations with a sun-like atmosphere model \citep{f01} using a kind of relaxation method \citep{r96}. Compare with \citet{zwd13}, the velocity viscous is replaced by a velocity damping term since velocity can not be dissipated effectively by former. The large velocity produced by the strong forced magnetic field of AR leads to a failed extrapolation. With source terms for gravity and velocity damping, the new MHD equations have the following form:
\begin{eqnarray}
\frac{\partial {\rho}}{\partial t} + \nabla\cdot (\rho {\mathbf{v}}) &=& 0,
\label{eq:mass} \\
\frac{\partial {\rho \mathbf{v}} }{\partial t} + \nabla\cdot\left[{\rho \mathbf{vv}+\left(p+\frac{\mathbf{BB}}{2}\right)\bf I-BB}\right] &=& \rho \mathbf{g}-\mu\rho\mathbf{v},
\label{eq:momentum} \\
\frac{\partial E}{\partial t} + \nabla\cdot \left[\left(E+p+\frac{\mathbf{B}\cdot \mathbf{B}}{2}\right)\bf v - B (B \cdot v)\right] &=&-\mu\rho v^{2},
\label{eq:inter energy} \\
\frac{\partial {\mathbf{B}}}{\partial t} - \nabla \times (\bf v \times B) &=& 0,
\label{eq:induction}
\end{eqnarray} where $\rho$, $\mathbf{v}$, $E$, $\mathbf{B}$ and $p$ are the mass density, velocity, total energy density, magnetic field and gas pressure, respectively. $g=274m\cdot s^{-2}$ is the surface gravitational acceleration. The frictional coefficient $\frac{1}{\mu}$ represents the timescale of the artificial velocity damping. The damping term is very important. Different forms of damping are applied to control the relaxation speed of the system by different relaxation method \citep{r96,bpm00,vkf07,jf12}. In this calculation, $\mu=0.1$.

Anther modification is the elimination of $\nabla \cdot \mathbf{B}$. In \citet{zwd13}, Powell's source terms \citep{prl99} are applied to prevent the accumulation of $\nabla \cdot \mathbf{B}$ by propagating the ``magnetic monopoles'' with flow. In this paper, the project scheme originally proposed by \citet{bb80} is further used to eliminate the $\nabla \cdot \mathbf{B}$ error. The Conjugate Gradient iterative method \citep{hs52} is used to solve the poisson equation in project scheme.

As claimed by \citet{zwd13}, it is important to choose a reasonable starting height of the atmosphere model as the initial condition to adapt the magnetogram. If too low, the heavy plasma is hard to be driven, which results too long time relaxation. If too high, the forced magnetic field is hard to be balanced with light plasma, which results failed extrapolation with severely distorted field lines. In this calculation, we choose a optimized starting height $z_{0}$ at which the ratio of gas pressure and average magnetic pressure (only include $B>200G$ area) on the magnetogram is equal to 1.

More details of the forced field extrapolation method can be found in \citet{zwd13}.

\section{Observational data}

\begin{figure}
\centering%
\includegraphics[width=10cm]{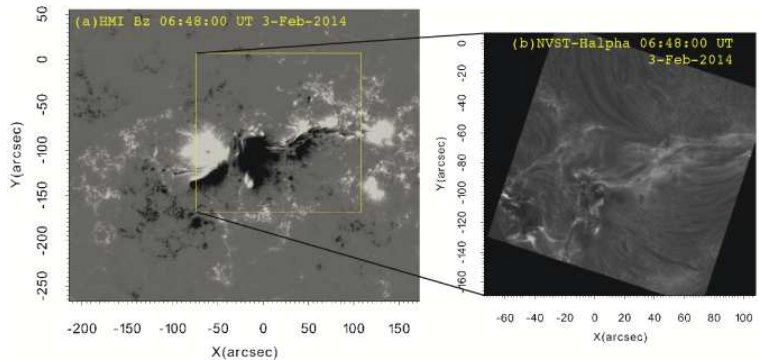}
\caption{Observations of AR 11967: (a) HMI line-of-sight magnetogram, (b) NVST-H$\alpha$ image shows the subregion outlined by square in (a).} \label{fig:observation}
\end{figure}

The NVST located at the northeast side of Fuxian Lake in China has three channels to observe the sun. The H$\alpha$ 6535$\AA$ channel with bandwidth of 0.25$\AA$ is used for observing chromosphere. The H$\alpha$ data adopted here cover part of the central area of AR11967 with a field of view of $151''\times151''$ and a pixel size of $0''.163$ at 06:48:00 UT on 2014 February 3. The HMI on board the SDO measures the vector magnetic field with pixel size of $0''.5$ at the same time. Fig.\ref{fig:observation} shows the observation of AR 11967 by HMI and NVST-H$\alpha$.

\section{Extrapolation and comparison}

We carry out computations in a box of $768\times640\times80$ grids with the same resolution as HMI magnetogram by three different algorithms: potential method (``Pot''), optimization method using grid refinement and preprocessed magnetogram implemented by \citet{w04} (``Wie''), forced field extrapolation (``Fce''). Unless otherwise stated, only the results below 10 grids (3.6 Mm) is considered.

\begin{figure}
\centering%
\includegraphics[width=14cm]{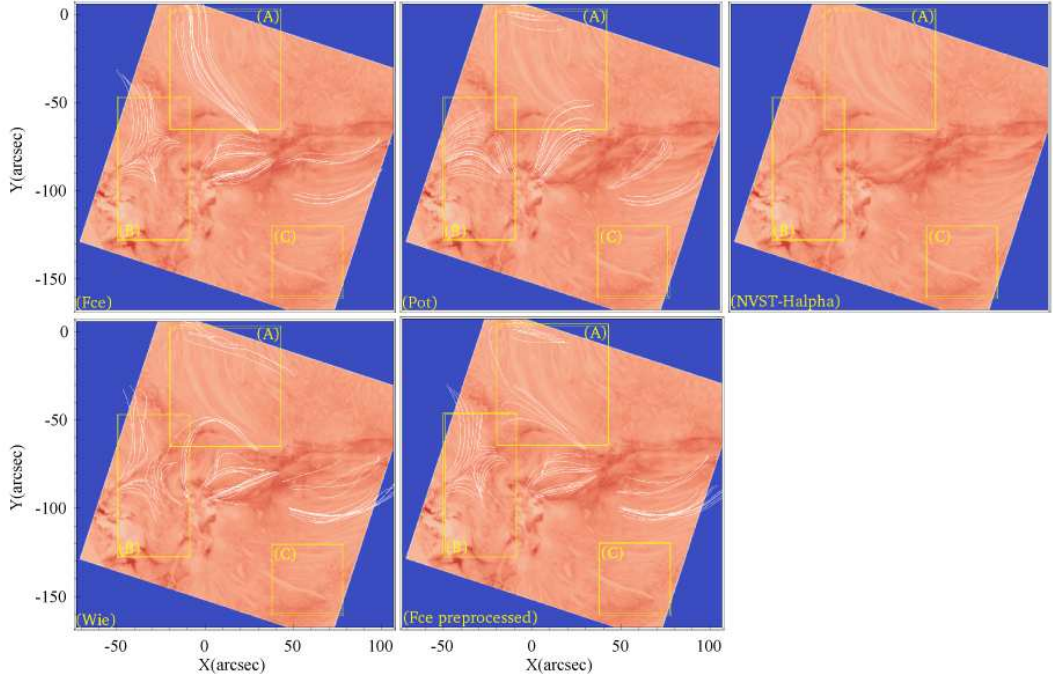}
\caption{Comparison of the NVST-H$\alpha$ image with field lines computed by ``Pot'', ``Wie'' and ``Fce'' models. ``Fce preprocessed'' is the ``Fce'' result with preprocessed magnetogram used as input. The field lines of each panels start from the same location.} \label{fig:comparenvst}
\end{figure}

The metrics describe the quality of the results are: current weighted average of $sin\theta$ (CWsin,\citet{wsr00}), unsigned mean of the absolute fractional flux ratio $<|f|>$ and magnetic energy. They are defined as:
\begin{eqnarray}
CWsin = \frac{\int_{V}J\sigma dV}{\int_{V}JdV},\quad \sigma = sin\theta = \frac{|\mathbf{J}\times \mathbf{B}|}{JB},
\label{eq:cwsin}
\end{eqnarray}
\begin{eqnarray}
<|f|>=\frac{1}{V}\int_{V}\frac{\nabla \cdot \mathbf{B}}{6B/\Delta x}dV,
\label{eq:divB}
\end{eqnarray}
\begin{eqnarray}
E_{B}=\int_{V}\frac{B^{2}}{8\pi}dV,\quad E_{pot}=\int_{V}\frac{B^{2}_{pot}}{8\pi}dV,\quad E_{free}=E_{B}-E_{pot},
\label{eq:magnetic-energy}
\end{eqnarray}
\begin{eqnarray}
lost\ E_{free}=(E_{free(fce)}-E_{free(model)})/E_{free(fce)},
\label{eq:lost-free}
\end{eqnarray} where $E_{B}$ and $E_{pot}$ are extrapolated magnetic field energy and potential energy, ``lost $E_{free}$'' is the relative free energy lost of a model compare with ``Fce'' result. Table \ref{tab:metric} contains metrics characterizing four regions: full, A, B and C(see outlined squares in figure \ref{fig:comparenvst} and \ref{fig:Bevolve}). The magnetic field is strong(weak) in square B(A and C). The CWsin metric in B is smallest, since strong field region is close to force-free state. In ``Fce'', the full region CWsin is 0.42 which is larger than ``Wie'' (0.13) and previously reported result of NLFFF model apply to active regions \citep{sdm08,dsb09,jf13}. Such result is expected because the raw magnetogram is used in ``Fce'', while ``Wie'' here use preprocessed magnetogram in which the net force and torque are eliminated. Table \ref{tab:metric} also shows ``Fce'' contains a little less free energy than ``Wie'' in the full domain. In square A, however, ``Fce'' contains much more free energy than ``Wie''.

In Figure \ref{fig:comparenvst}, we compare the extrapolated field lines with NVST-H$\alpha$ fibrils in chromosphere. As expected, non-potential extrapolations show better results than ``Pot''. To see clearly, we outline squares A and B which is the same subregion in figure \ref{fig:Bevolve}.

Square B shows a typical X-shape structure where reconnection took place. The X-shape identified by H$\alpha$ fibrils indicate the field line orientation. Both ``Fce'' and ``Wie'' results are in good agreement with fibrils which located at the lower and right side of the null point. At the left side, ``Fce'' field lines show more real bend than ``Wie''. We cannot determine which results are better at the upper side. Generally, ``Fce'' results reveal a closer match to the observed X-shape structure.

In square A, we see very different field orientation. ``Fce'' field lines are closely resemble the H$\alpha$ fibrils while ``Wie'' shows obvious misalignment which is similar to ``Pot'' results. The metrics in table \ref{tab:metric} demonstrate square A is a very non-potential ($E_{B}/E_{pot}=2.06$) region. ``Fce'' shows the ability in recovering the magnetic field in non-potential region.

\begin{figure}
\centering%
\includegraphics[width=10cm]{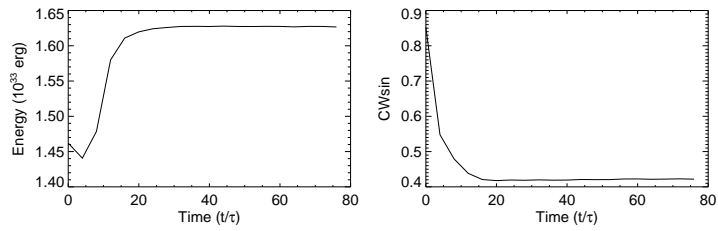}
\caption{Temporal evolution of magnetic energy (left) and CWsin (right).} \label{fig:convergence}
\end{figure}

\begin{figure}
\centering%
\includegraphics[width=14cm]{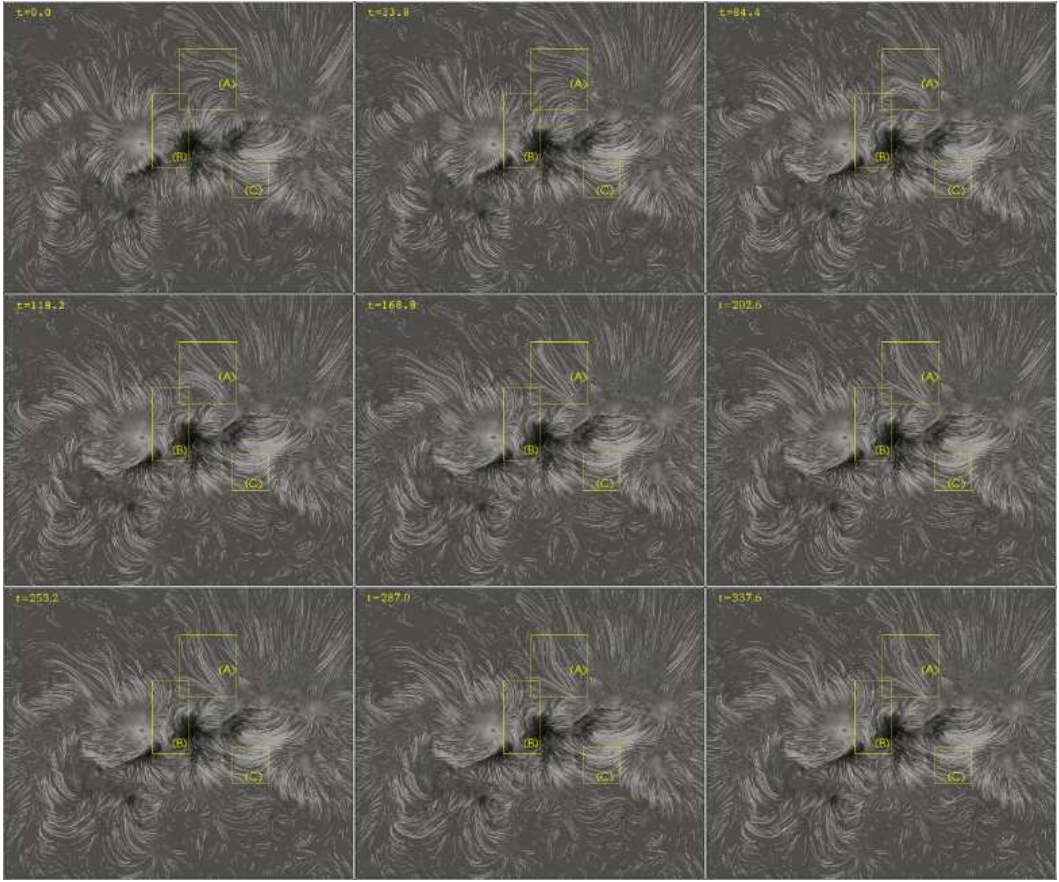}
\caption{Temporal evolution of magnetic field lines from T=0 to T=337.6$\approx 70\tau  $ against the map of $B_{z}$.} \label{fig:Bevolve}
\end{figure}

Figure \ref{fig:convergence} shows rapid convergence of magnetic energy and the angle between $\mathbf{B}$ and $\mathbf{J}$ during relaxation. The energy and current weighted sin metrics reach final values after times t=20 and 30 $\tau$ ($\tau$ is the fast wave cross time), respectively. To further check the convergence, we plot temporal evolution of field lines (see figure \ref{fig:Bevolve}) during relaxation process. We note the field lines at many subregions (especially A and B) slowly changing the orientation, which indicate the injection or transfer of energy. The magnetic field reach the stable final state after about $t=168 \approx 35\tau$. The final state shows obviously non-potential configuration compare with initial state.

\section{Discussion}

\subsection{Preprocess}
It has been very popular to preprocess the magnetogram \citep{fsv07,sdm08,twi09,dsb09,gdc13,cdz14,jwf14,wld15}, which makes it more consistent with NLFFF. \citet{mds08} showsed preprocessing improve the metrics quantify the agreement of NLFFF solution and reference model field. \citet{sdm08} applied 14 NLFFF models with 4 different codes and a variety of boundary conditions on AR10930, they obtain best results for preprocessed data.

\begin{figure}
\centering%
\includegraphics[width=10cm]{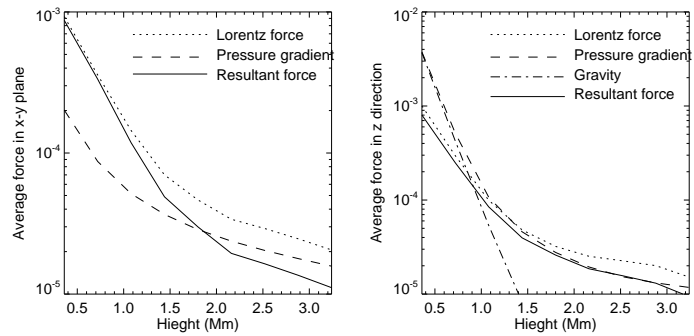}
\caption{Horizontal average of different forces, $\frac{1}{N}\sum (f_{x}^{2}+f_{y}^{2})^{1/2}$ (left), $\frac{1}{N}\sum |f_{z}|$ (rigtht).} \label{fig:force}
\end{figure}

For forced field extrapolation used here, however, the force in the magnetogram is very important in driving the MHD system to a forced equilibrium. This can be indicated in figure \ref{fig:force}. In x-y plane, the average of resultant force is smaller than lorentz force at any height, which means lorentz force is partly canceled out by pressure gradient. In z direction, the situation is more complicated because the gravity is included. Due to the heavy plasma around the bottom boundary, the pressure gradient and gravity are dominant forces. The resultant force in z direction remain smaller than lorentz force at any height. Moreover, we repeat our relaxation with the same initial and boundary condition but a preprocessed magnetogram used as input. Smoothing of the magnetic field by preprocessing procedure can be seen clearly in figure \ref{fig:raw-pre-magnetogram}, which leads to more force free results as expected (see ``Fce preprocessed'' metrics in table \ref{tab:metric}). However, the field lines of ``Fce preprocessed'' show obvious misalignment with fibrils in square A (see figure \ref{fig:comparenvst}). And the ``Fce preprocessed'' result lost 13\% of free energy compare with ``Fce'' result in the whole region (see lost $E_{free}$ metric in table \ref{tab:metric}).

\subsection{The height of forced field}
\begin{figure}
\centering%
\includegraphics[width=6cm]{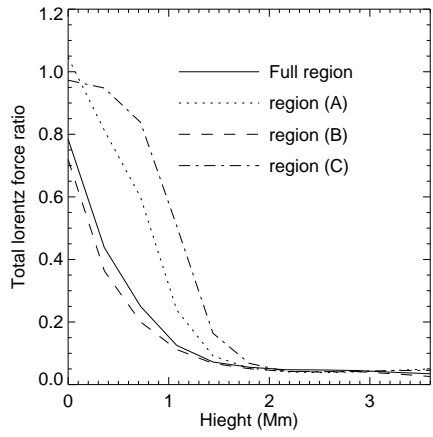}
\caption{Horizontal average of unsigned lorentz force, measured relative to averaged magnetic pressure gradient with height.} \label{fig:force-ratio}
\end{figure}

\begin{figure}
\centering%
\includegraphics[width=12cm]{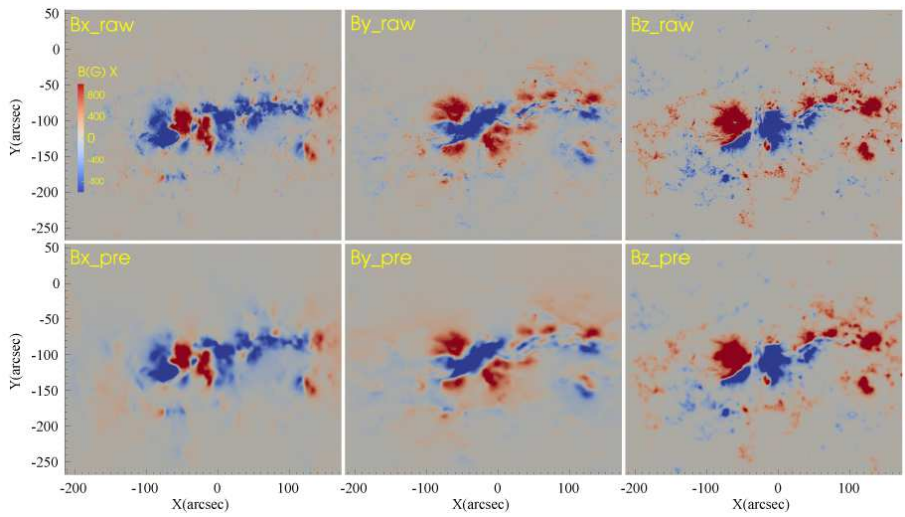}
\caption{Raw and preprocessed magnetogram of AR11967.} \label{fig:raw-pre-magnetogram}
\end{figure}

To derive the height under which lorentz force cannot be ignored, we plot the average, unsigned total lorentz force, divided by average magnetic pressure gradient at each height (see figure \ref{fig:force-ratio}). The lorentz force ratio in the strong magnetic field volume is larger than 0.1 below $1400km$, while the height reach $1800km$ in weak field. The forced region is much higher than 400km that is computed by \citet{mjm95}. Such forced chromosphere state is also found in recent radiation-MHD simulation by \citet{lcr15}.

\subsection{Why A is a special region?}
From above comparisons, we find the magnetic field computed by ``Fce'' and ``Wie'' differ most in square A. What is the reason?

\begin{figure}
\centering%
\includegraphics[width=10cm]{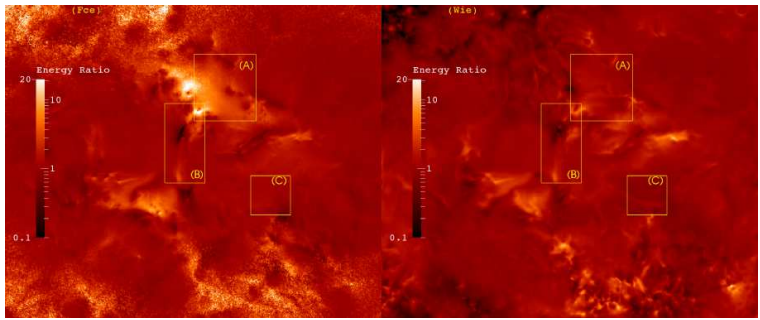}
\caption{Magnetic energy integrated along z direction below 10 grids divided by integrated potential energy.} \label{fig:energy-ratio}
\end{figure}

We first note A is a weak field region. However, it is not the sufficient condition because the two models show the similar field lines in C where the magnetic field is also weak. Then we note the energy ratio metric $E_{B}/E_{pot}$ which describe the deviation of the extrapolation from potential field. In square A, this metric is far larger than other region. Furthermore, we calculate the metric everywhere and show it in figure \ref{fig:energy-ratio}. We can see square A locates at the large energy ratio region of ``Fce'', while the ratio is small in square C. However, we do not see large energy ratio in square A from ``Wie''. So we believe square A contains large percent of free energy, and forced field extrapolation can reconstruct magnetic field in this region more successfully than NLFFF model. The snow appear near the upper and bottom boundary of ``Fce'' stern from the noise in the raw magnetogram. The ``Fce'' with preprocessed boundary eliminates the snow (not show here). The free energy lost metric (see table \ref{tab:metric}) shows ``Wie'' result contain 17\% more free energy than ``Fce'' result. In square A, however, ``Wie'' lost 69\% of the free energy.

It must be pointed out that due to a highly inclined geosynchronous orbit of SDO, the relative velocity between HMI instrument and the sun ranges $\pm$ 3km/s \citep{hlh14}. This leads to the temporal and spatial variations of inverted magnetic field over a period of 12 hr. The systematic errors in magnetic field data contaminate the outputs \citep{sal15}. The unsigned fluxes in Fig.3 of \citet{cz13} show obvious dip near the midnight on 14-Feb, 15-Feb and 16-Fe. The profile of energy flux in the bottom of Fig.2 of \citet{v15} shows clear oscillation over timescale of 12 hr. The free energy of the extrapolated magnetic field is also affected by this systematic errors.

\section{Conclusion}
We have applied forced field extrapolation method to model the chromospheric magnetic field above the AR11967 with HMI magnetogram. We conclude that: (1) the relaxation process is converged by checking the temporal evolution of CWsin, magnetic energy and magnetic field lines; (2) ``Fce'' can recover the field lines successfully in chromosphere, while `Wie'' may failed in the region contains large percent of free energy; (3) the raw magnetogram without preprocessing is required for ``Fce''; (4) The ``Fce'' results show the forced region can reach the height of $1400-1800km$ which is much higher than estimated before.

\begin{deluxetable}{ccccccc}
\tabletypesize{\scriptsize}
\tablecaption{Extrapolation metrics for AR11967. \label{tab:metric}}
\tablewidth{0pt}
\tablehead{\colhead{Model} & \colhead{CWsin} &\colhead{$<|f|>\times(10^{4})$} & \colhead{$E_{B}$} & \colhead{$E_{pot}$} & \colhead{$E_{B}/E_{pot}$} & \colhead{lost $E_{free}$(\%)}}
\startdata
Fce  &     &        &      &      &  & \\
Full &0.42 &2.58 &16.27 &14.62 &1.11 &- \\
A    &0.65 &2.21 &0.111 &0.054 &2.06 &- \\
B    &0.23 &1.47 &4.02  &3.69  &1.09 &- \\
C    &0.65 &4.52 &0.059 &0.054 &1.10 &- \\
\tableline
Wie  &     &        &      &      &      &\\
Full &0.13 &11.7 &16.52 &14.59 &1.13 &-17.1\\
A    &0.11 &10.1 &0.070 &0.050 &1.39 & 65.1\\
B    &0.09 &10.2 &4.10  &3.71  &1.11 &-17.0\\
C    &0.10 &8.86 &0.043 &0.054 &0.79 &  3.0\\
\tableline
Fce preprocessed  &     &        &      &      &  &\\
Full &0.34 &1.56 &15.80 &14.37 &1.10 & 13.1\\
A    &0.56 &1.22 &0.101 &0.050 &2.02 & 10.4\\
B    &0.21 &0.82 &3.93  &3.64  &1.08 & 15.3\\
C    &0.53 &2.40 &0.061 &0.053 &1.15 &-38.8\\
\enddata
\tablecomments{Energy unit is $10^{33}$ erg.}
\end{deluxetable}


\section*{Acknowledgement}\footnotesize

This work is jointly supported by National Natural Science Foundation of China (NSFC) through grants 11403044, 11473040 and 11273031. The data used are courtesy of the NVST and SDO science teams.

\bibliographystyle{apj}\footnotesize
\bibliography{ms}\footnotesize

\end{document}